# Differenciated Bandwidth Allocation in P2P Layered Streaming

Abbas Bradai, Toufik Ahmed
CNRS-LaBRI University of Bordeaux-1
351, Cours de la libération. Talence, 33405
{bradai, tad} @labri.fr

*Abstract*— **There is an increasing demand for P2P streaming in particular for layered video. In this category of applications, the stream is composed of hierarchically encoded sub-streams layers namely the base layer and enhancements layers. We consider a scenario where the receiver peer uses the pull-based approach to adjust the video quality level to their capability by subscribing to different number of layers. We note that higher layers received without their corresponding lower layers are considered as useless and cannot be played, consequently the throughput of the system will drastically degrade. To avoid this situation, we propose an economical model based on auction mechanisms to optimize the allocation of sender peers' upload bandwidth. The upstream peers organize auctions to "sell" theirs items (links' bandwidth) according to bids submitted by the downstream peers taking into consideration the peers priorities and the requested layers importance. The ultimate goal is to satisfy the quality level requirement for each peer, while reducing the overall streaming cost. Through theoretical study and performance evaluation we show the effectiveness of our model in terms of users and network's utility.**

## I. INTRODUCTION

Peer-to-peer (P2P) networks are getting increasingly popular for streaming video over the Internet. Due to peer dynamics, single-layer stream can neither match the overlay capacity changing, nor meet heterogeneous peer requirements. Layered streaming, such as Scalable Video Coding (SVC), provides a convenient way to perform video quality adaptation to adjust to the changing network conditions and receiver preferences. A layered streaming consists of a base layer and multiple enhancement layers. Receivers can adjust the video quality level to their capability by subscribing to different number of layers using pulling distribution approach. In a P2P network, it is natural to request the layer from different peers (upstream peers). Thus, each upstream peer shares its upload bandwidth among different peers to serve different layers. How to resolve bandwidth conflicts among peers in order to maximize benefits of both upstream and downstream peers while respecting the layers importance, their dependencies and the peers' priorities is highly challenging in P2P networks.

For P2P streaming applications, bandwidth allocation is an important factor because of its direct bearing on high quality and lower latency performance. Kumar et al. [1] build a fluid model to study the impact of peers' upload bandwidth and conclude conditions for universal streaming for churnless systems. Zhang et al. [2] study the bandwidth influence on chunk (data block) scheduling algorithms using extensive packet-based simulations. They prove that random chunk scheduling can achieve near-optimal streaming quality if the overall upload bandwidth is at least 1.2 times of needed bandwidth. Furthermore, measurement studies and implementations [3] [4] show also that bandwidth has a big impact on streaming quality for P2P streaming systems. Recently, researchers have studied the bandwidth allocation for improving streaming quality in more challenging P2P networks such as multi-overlay, multi-sources and multi-Swarm P2P streaming systems. Wu et al. [5] study the bandwidth contest among coexisting overlays and propose a solution based on auction. Liang and Liu [6] study the optimal bandwidth sharing in multiple video conferencing swarms systems. They dynamically share a pool of helpers between swarms to address the bandwidth shortage intra and inter-swarms. However, none of these works have taken into consideration the layered streaming properties, mostly the layers dependency.

Many recent works, such as [7] [8], leverage the characteristics of SVC and P2P networks and propose adaptive video streaming mechanisms. In [7], authors propose an optimization technique based on harmony search algorithm in order to increase the delivery ratio for the most important layers, while reducing the overhead and ensuring load balancing in the overlay. Authors in [8] propose taxation based P2P layered streaming designs, including layer subscription policy, chunk scheduling strategy and mesh topology adaptation. No one of these works has tackled the upstream peer bandwidth allocation problem in layered P2P systems. Indeed, a little literature has studied bandwidth conflicts for layered streaming. To our knowledge, the most closely related work is presented by Wu et al [5]. They coordinate multiple streams as an auction game, where each peer participates in media distribution by bidding for and selling bandwidth. Their strategy is not compatible with scalable streams as each stream is considered as an isolated stream without any relationship with any other streams. In contrast, there exists inherent content priority among layers. Moreover, authors only consider a scenario where the upload bandwidth sum in the network is always sufficient to support all the peers' requirement in all the overlays.

In this paper, we look for resolving bandwidth conflicts in P2P layered streaming. Our main contributions are as follows: Firstly, we model the bandwidth allocation process in P2P layered streaming as a series of auction games [9] in which peers bid for and sell the upload bandwidth to maximize their benefits. In order to resolve the problem of layers dependency, we set up auction game to allocate bandwidth for each layer. We start by allocating the bandwidth for the lower layers then for the upper's ones. It is important to note that in our model we take into consideration the requirements of peers in terms of video quality level as well as their priorities (QoS level of the peer) using a system of budget allocation for peers.

The rest of the paper is organized as follows. In section II we describe in detail the proposed auction mechanism for bandwidth allocation and the related theoretical study. In Section III, we present some illustrative simulation results. Finally, section IV provides the conclusion and future work.

## II. THE PROPOSED ALLOCATION MECHANISM

In this section, we develop the proposed mechanism for bandwidth allocation problem in P2P layered streaming using microeconomic theory, namely the auction mechanism.

### 1. Network model and assumptions

We consider an overlay network composed of *n* peers. Among them *m* upstream peers sharing their upload bandwidth to serve *l* downstream peers, relayed by a set of application-layer links $m_{i,j}$ (link between downstream peer *i* and its upstream peer *j*). So, the topology of the overlay can be modeled as a directed weighted graph *G*= (*S, L, K*) where *S* denotes the set of upstream peers (called in some architecture Seeders), *L* the set of downstream peers (called in some architecture Leechers) and *K* the set of links.

Suppose the layered stream is encoded into M layers at the source peer $\{l_0, l_1 \ldots l_M\}$, with $l_0$ representing the base layer and $l_1 \ldots l_M$ representing the enhancement layers, respectively. We assume that each video layer $l_i$ is distributed with a transmission rate of $B_i$. Thus, each peer can subscribe to a particular video layer depending on its download capacity and other parameters such as its processing capability, preferences, etc. In this paper, we assume that the peers decide their quality level only depending on their available download bandwidth. Besides, we assume that the quality level does not vary frequently, by applying a certain smoothing function as described in [10].

To represent a practical network setting, we limit the upload and download capacity of any peer ∈ G by $u_i$ and $b_i$. Therefore, each peer can only provide limited service for its downstream peers, and make a limited layer subscription as well. We assume also that the downstream peers have different levels of priorities $P=\{pr_1, pr_2, \ldots, pr_q\}$ where $pr_1 > pr_2 > \ldots > pr_q$. So, for each downstream peer $i \in L$ is assigned a level of priority $pr_i \in P$. This can be mapped in real world P2P systems to an incentive mechanism where peers contributing in the overlay (sharing more upload bandwidth for example) are promoted to upper priority classes, and peers less contributing are demoted to lower priority classes.

### 2. Bandwidth allocation model

The process of bandwidth allocation is modeled as a set of dynamic auction organized by upstream peers in order to give rise to competition on its upload bandwidth $u_j$. The players in this auction game are the downstream peers. Indeed, each downstream peer, having an initial budget $T_i$, submits bids to all its downstream peers in order to purchase bandwidth. Consequently, each downstream peer can participate in different independent parallel auction games $A_j$, organized by its upstream peers.

In order to guarantee a minimum quality level for all downstream peers, and respect the layers dependency, every upstream starts by allocating bandwidth for the base layer, than for the enhancement layers in an ascending manner. Concretely, every upstream peer organizes an auction to distribute bandwidth needed for the base layer. Then, if there is still remaining bandwidth, it organizes another auction to sell bandwidth for the first enhancement layer, the second enhancement layer, and so on. In the receivers' side, the peers participate in auctions depending on the quality level that they decide. A peer which has decided a quality level 2 will participate in all auctions organized by its upstream peers to distribute bandwidth for base layer ($l_0$), for the first enhancement layer ($l_1$) and for the second enhancement layer ($l_2$). We note that an upstream peer doesn't start to allocate bandwidth for an upper layer, until there is no request for the current layer from their downstream peers (i.e. the downstream peers' requests are satisfied or their budget is exhausted).

A set of auction games, to distribute bandwidth for certain layer, is illustrated in Figure 1. In this figure, three parallel auctions are organized by upstream peers S1, S2 and S3 in order to allocate their upload bandwidths $u_1$, $u_2$ and $u_3$ respectively for certain layer $l_k$. The downstream peer $L_1$, connected to upstream peers $S_1$, $S_2$ and $S_3$, participates in auctions $A_1$, $A_2$ and $A_3$ organized by the three upstream peers. While the downstream peer $L_4$ participates only in auction $A_3$ organized by $S_3$ (since $L_4$ is connected only to $S_3$).

Let $b_{i,j}^k$ be the bandwidth requested by the downstream peer *i* to its upstream peer *j* to acquire bandwidth for layer *k* and $p_{i,j}^k$ be the unit price that the downstream peer is willing to pay for that bandwidth. The bid of the downstream peer *i* can be expressed by the pair $B_{i,j}^k = (b_{i,j}^k, p_{i,j}^k)$

After modeling the bandwidth allocation problem in P2P layered streaming as a set of auction games, where items to sell are the upload bandwidth, and where the sellers are the upstream peers and the buyers are the downstream peers, we discuss the allocation strategies of the upstream peers and the bidding strategies followed by the downstream peers.

### A. Upstrem peer's side : bandwidth allocation strategy

As mentioned before, the upstream peer starts allocating bandwidth foremost for the lowers layers then the upper layers. It executes the algorithm presented in Table 1.

$l_k = l_0$
**While** $u_i > 0$ and $l_k \leq l_M^i$
  Auction ($l_k$)
  $u_i = u_i - u_i^k$
  $l_k = l_{k+1}$
**End while**

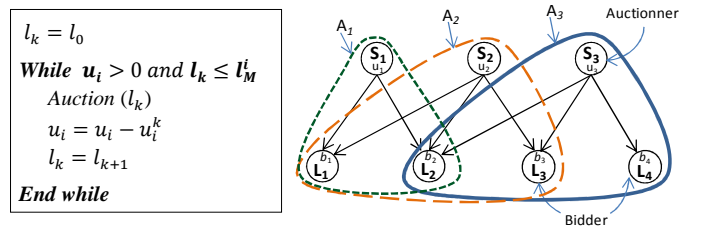

Table 1: Allocation strategy        Figure 1: Auctions example

Where $u_i^k$ represents the total bandwidth allocated to the layer *k* for all peer *i*' downstream peers and $l_M^i$ is the maximum layer available in the peer *i*.

In the following, we detail the strategy of the upstream peer within the auction game to allocate bandwidth for a layer *k*. In auction $A_j^k$, organized by the peer *j* to allocate bandwidth for the layer *k*, the seller *j* aims to maximize its revenue by selling its bandwidth at the best price. Given the downstream

peers' bids $B_{i,j}^k = (b_{i,j}^k, p_{i,j}^k)$, the upstream peer $j$ aims to maximize:

$$\max \sum_{i \in L_j} (b_{i,j}^k p_{i,j}^k) \quad (1)$$

Subject to: $\quad \sum_{i \in L_j} b_{i,j}^k \leq u_j \quad (2)$

Where $L_j$ denotes the set of downstream peers connected to the upstream peer $j$.

In order to maximize its revenue, the upstream peer adopts the best offer auction strategy: it starts first by serving the downstream peer, willing to pay the highest price. Once it is served and if there is still remaining bandwidth, the downstream peer proposing the second highest price will be served and so on.

The allocation strategy of the bandwidth for a layer $k$ is performed in many rounds as shown in Table 2.

1) Receive bids from downstream peers
2) Allocate bandwidth to downstream peer willing to pay the highest price
3) While there is still remaining bandwidth serve the downstream peer willing to pay the next highest price
4) Notify the allocated bandwidth to all the downstream peers involved in the auction
5) Notify the allocated bandwidth to all downstream peers involved in the auction
6) Receive new bids from downstream peers (having sufficient budget) whose bandwidth request is not satisfied. Go to 2

Table 2: Bandwidth allocation strategy

*B. Downstream side: bidding strategy*

As presented earlier, the bidding of a downstream peer $i$ in auction organized by an upstream peer $j$ to sell bandwidth for layer $k$ is the pair $B_{i,j}^k = (b_{i,j}^k, p_{i,j}^k)$. The question to deal with here is what strategy should be followed by the downstream peer to set $b_{i,j}^k$ and $p_{i,j}^k$? In other words how much bandwidth downstream peer requests to each of its upstream peer and at what price?

The ultimate goal of the downstream peer is to minimize the bidding cost as well as the streaming cost. The bidding cost can be mapped in real world's auction to the items' purchase price which express the competition degree on these items. Indeed, we believe that requesting stream from less loaded upstream peers, allows reducing the delay, since it reduces the congestion in the concerned peers. In addition it allows a good load balancing of the stream through the overlay, which can be benefic in the case of peers churn. The streaming cost can be seen as the transport cost of purchased items. In the context of P2P streaming, reducing the streaming cost is equivalent to get the stream from best links. That means links with lowest delay, lowest bit error rate, etc. So, the goal of the downstream peer is to get bandwidth from less loaded upstream peers (low price) and via best links (low streaming cost).

As introduced before, the downstream peer starts first by requesting bandwidth for the lower layers, the enhancement layers incrementally, by joining the corresponding auction organized by the upstream peers in this order. This strategy allows requesting primarily the bandwidth for the lower layers from the best links and then the enhancement layers from the other links. Hence, lower layers have more chance to be received by the downstream peers, and consequently more chance to decode the layered stream properly. In the opposite, if the upper layers are promoted, the decoding of the stream could not be possible in the case of the corresponding lower layers are missing and by consequence, the throughput of the system will degrade.

Formally, in each auction games organized by an upstream peer $j$ to allocate bandwidth for the layer $k$, the downstream peer aims to minimize the bidding cost. That means:

$\min \sum_{j \in s_i} b_{i,j}^k p_{i,j}^k \quad$ Subject to $\quad \sum_{j \in s_i} b_{i,j}^k \geq B_k \quad (3, 4)$

Where $S_i$ denotes the set of upstream peers of the peer $i$ and $B_k$ denotes the transmission rate of layer $k$.

In addition to the bidding cost, the downstream peer aims also to minimize the streaming cost from each of its upstream peer $j$, denoted as $E_{i,j}(b_{i,j}^k)$. Therefore, the bidding strategy of downstream peer $i$, in each auction game $A_i^k$ to acquire bandwidth for layer $k$, can be seen as an optimization problem of the overall cost:

$$\min \sum_{j \in s_i} (b_{i,j}^k p_{i,j}^k + E_{i,j}(b_{i,j}^k)) \quad (5)$$

Subject to $\quad \sum_{j \in s_i} b_{i,j}^k \geq B_k$ and $b_{i,j}^k \geq 0 \quad (6, 7)$

In practice, we consider the streaming cost function $E_{i,j}$ as non-decreasing function depending on $b_{i,j}^k$, strictly convex and twice derivable.

In the following we present the bidding strategy of the peer to set the requested bandwidth ($b_{i,j}^k$) and the bidding unit price ($p_{i,j}^k$).

a) *Peer's strategy to set the requested bandwidth* ($b_{i,j}^k$)

Given the bid price $p_j^k$ in an auction organized by the upstream peer $j$ to allocate bandwidth for layer $k$, the downstream peer $i$ aims to optimize the overall cost by adjusting the requested bandwidth $b_{i,j}^k$ from each downstream peer. So, the goal of the downstream peer is to minimize the global marginal cost $M_i$ defined as the change in total cost that arises when the quantity produced changes by one unit [11].

Let $c_i$ be the overall cost at the downstream peer $i$, i.e.

$$c_i = \sum_{j \in s_i} (b_{i,j}^k p_{i,j}^k + E_{i,j}(b_{i,j}^k)) \quad (8)$$

The corresponding marginal cost is:

$$M_i = \frac{dc_i}{db_{i,j}^k} = \sum_{j \in s_i} p_{i,j}^k + \sum_{j \in s_i} \frac{dE_{i,j}(b_{i,j}^k)}{db_{i,j}^k} \quad (9)$$

Since the streaming cost function $E_{i,j}$ is strictly convex, the second derivative $\frac{dM_i}{db_{i,j}^k} = \frac{d''E_{i,j}(b_{i,j}^k)}{db_{i,j}^k}$ is strictly positive. Consequently, the marginal cost $M_i$ increases with the increase of the bandwidth request $b_{i,j}^k$. To solve efficiently this optimization problem we consider the water filling algorithm [12]. To set the bandwidth quantity (to request from an upstream peer $j$), the downstream peer $i$ - applying the water filling algorithm - set $b_{i,j}^k$ to 0 for all $j \in S_i$, then it identifies the upstream peer $j_0$ having the lowest marginal cost $M_{i,j0}$ and increases its demand $b_{i,j_0}^k$ until the marginal cost becomes equal to the next highest marginal cost $M_{i,j1}$, corresponding to the upstream peer $j_1$. The downstream peer $i$ increases then fairly $b_{i,j_0}^k$ and $b_{i,j_1}^k$ until their corresponding marginal cost $M_{i,j0}$ and $M_{i,j1}$ meet the next highest marginal cost $M_{i,j2}$ corresponding to the upstream peer $j_2$, and so on. The downstream peer carries out this mechanism with all its upstream peers until it obtains the bandwidth that it requests for the layer $k$ ($B_k$).

*b) Peer's Downstream peer's strategy to set the bidding unit price ($P_{i,j}^k$)*

After defining the bandwidth request strategy of the downstream peer, the next question to deal with is how the downstream peer set the unit price that it will announce to the upstream peer $j$?

In the bootstrap stage of each auction, the downstream peer is provided with an initial budget $T_i^k$ for each layer $k$, which it spends to acquire bandwidth for the layer $k$. This budget is relative to the priority of the peer. The peer with higher priority receives larger budget, and vice versa. The budget $T_i^k$ of the downstream peer $i$ is defined by the formula:

$$T_i^k = \tilde{p}_i^k B_k \quad (10)$$

Where $\tilde{p}_i^k$ denotes the reference unit price assigned to the downstream peer $i$.

When the downstream peer $i$ joins the auction organized by a upstream peer $j$, first, it sets its price bid to one unit (i.e. $p_{i,j}^k = 1$). Using the water filling algorithm described earlier, it computes the optimal quantity of bandwidth $b_{i,j}^k$ to request from each upstream peer $j$, and submits bids consequently. After the upstream peers allocate their upload bandwidth using the strategy described above, it proposes the bandwidth $a_{i,j}^k$ to the corresponding downstream peers. On receiving the proposed bandwidth, the downstream peer determines its behavior in the next round of auctions, using the following algorithm:

1) Receive bandwidth allocation and current prices from upstream peer.
2) For each upstream peer:
   If requested bandwidth $b_{i,j}^k$ from an upstream peer $j$ is not satisfied:
   a. Increase the price $p_{i,j}^k$ by one unite within the reference price $\tilde{p}_i^k$
   b. Using the water filling algorithm, decide the quantity of bandwidth $b_{i,j}^k$ to request from $j$
3) Send new bids ($b_{i,j}^k, p_{i,j}^k$) to the upstream peer

Table 3: Bidding algorithm

It is clear in this algorithm that the reference price $\tilde{p}_i^k$ assigned to the downstream peer allows differentiating the downstream peers in accordance with their priorities.

*C. Convergence to Nash equilibrium*

Our mechanism for bandwidth allocation can be modeled as a non-cooperative game where the players are the set of downstream peers L, the strategies are the set of bids ($b_{i,j}^k, p_{i,j}^k$) and the cost function of a player $i$ is the overall streaming and bidding cost $C_i$. Formally, we consider the finite game $\Gamma = <$L, D, C$>$ where:
- L denotes the set of players (downstream peers)
- D denotes the set of strategies, i.e. D = (D$_1$, D$_2$ … D$_i$), where D$_i$ = ($B_{i,1}^k, B_{i,2}^k …B_{i,j}^k$) is the tuple of bids submitted by the player to its upstream peers.
- C denotes the set of costs: C = ($c_1, c_2 … c_i$), where $c_i$ is the overall cost at the player $i$ as defined in (8).

*Theorem*

*The auction game for bandwidth allocation in layered P2P leads to a Nash equilibrium*

*Proof*

Due to lack of space, details of the proof are omitted. The basic idea behind is to prove the Nash equilibrium of each auction $A_i^k$, and then derive the Nash equilibrium of the whole system. The proof of the Nash equilibrium in $A_i^k$ can be reduced to the proof of the existence of fixed point for the transfer function describing the evolution of the auction states from an auction round to another. After that we can conclude that our system presents a Nash equilibrium point, which is the set of Nash equilibrium points of all the auctions $A_i^k$.

### III. PERFORMANCE EVALUATION

In this section we describe the performance evaluation of the proposed bandwidth allocation mechanism for different QoS levels using Simulink-Matlab simulations [14].

*A. Simulation set up*

The performance of our system is carried for mesh based P2P network of different size to measure the performance of our mechanism in terms of:
- The delivery ratio of each layer
- Useless chunks ratio defined as the ratio of number of chunks received without their corresponding lower layers
- Downstream peers' average streaming cost
- The efficiency of peers priority distribution over the network

We generate diverse topologies of different sizes and different upstream peers' connectivity degree, defined as the number of downstream peers connected to the same upstream peer. Each network includes three classes of downstream peers: $Q_1$, $Q_2$ and $Q_3$ with priorities $pr_1$, $pr_2$ and $pr_3$ respectively.

The upload bandwidth of each upstream peer varies from 256 kbps to 2 Mbps and it is uniformly distributed throughout the network, while the download bandwidth of each downstream peer varies between 256 kbps and 1 Mbps.

The stream is subdivided to 6 layers. The bitrate of base layer is 200kbps, while the enhancement layer bitrate is 100kbps.

We use the following function to measure the streaming cost:

$E_{i,j}(b_{i,j}^k) = \frac{b_{i,j}^k}{x_{i,j} - b_{i,j}^k}$ where $x_{i,j}$ is the available bandwidth

between the downstream peer $i$ and its upstream peer $j$. This function express the ratio of the peer's $i$ requested bandwidth from the peer $j$, to the remained free bandwidth in the link $m_{i,j}$, i.e. the utilization ratio of the link. Link having low utilization ratio, is considered as good link because it presents low delay and low bit error rate since the intermediate routers' queues are less occupied. We note that we choose this streaming cost function as an example to perform our simulation. Any other function to evaluate the streaming cost in a link, satisfying the conditions of convexity and derivability mentioned in II.B, can be used.

*B. Results and discussions*

Due to space constraint, we present in this paragraph only some results that we obtained.

Figure 2 shows the average delivery ration of the different layers in an overlay of 500 nodes, while varying the upload capacity of the peers. The common observation for the delivery ratio in the different layers is that it increases with the increase of the upload capacity of the peers. In addition we observe that the delivery ratio of the lower layers is always higher than in the higher layer. This confirms our strategy to allocate the lower layers first than the upper ones. That allows the lower layers to get the best links, and consequently high delivery ratio

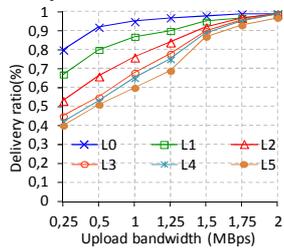

Figure 2: Delivery ratio Vs average available upload bandwidth

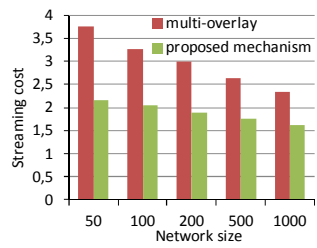

Figure 3: streaming cost Vs. network size

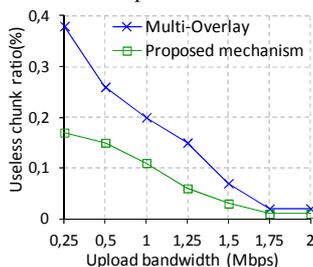

Figure 4: Useless chunk ratio Vs average available upload bandwidth

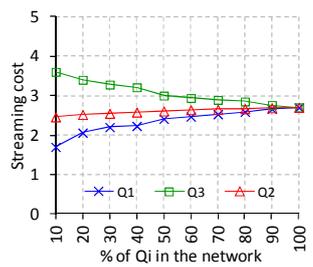

Figure 5: streaming cost Vs. network configuration

In Figure 3 we studied the average streaming cost in our system compared to multi-overlay auction mechanism, adapted to the multilayer. We perform the simulation in diverse topologies with different network size. In each network, 10% of peers belong to Q1, 30% to Q2 and 60% to Q3. It is observed that the average streaming cost decreases with the increase of the network size both in our solution as well as in the multi-overlay one. This can be explained by the increase of the number of links in the network, and consequently discharging the old ones. Nevertheless, in all configurations, the streaming cost is reduced by at least 25% in our solution compared to multi-overlay solution.

In Figure 4 we compare the performance of our mechanism with the multi-overlay one in terms of useless chunk ratio. We observe that the useless chunks are relatively high in the case of low average upload bandwidth, mostly in the case of the multi-overlay solution since it does not take into consideration the layers importance. With the increase of the available upload bandwidth, the useless chunk ratio decrease, because of the increasing availability of the bandwidth to require the different layers. Our mechanism starts by allocate bandwidth to lower layers first, consequently these layers are received through the best link, which enhance the delivery ratio of lower layers, consequently reduce the useless chunk ratio comparing to the other solution.

Figure 5 illustrates the evolution of the average streaming cost experienced by peers, by varying the size of each class of peers. The network is composed of 500 peers.

First of all, we observe that the streaming cost in downstream peers of $Q_1$ is smaller than in downstream peers of $Q_2$ which is smaller than in downstream peers of $Q_3$. This shows that the bandwidth allocation mechanism in our system respects the priorities of peers.

We observe also that with the increase in downstream peers of $Q_1$, the average streaming cost experienced by peers of this class increases. This is due to the increase in the competition on the good quality links (links with low streaming cost). As a result, more and more downstream peers of $Q_1$ get connected with links having higher streaming cost. On the opposite, the streaming cost of downstream peers with $Q_3$ decreases with the increase of their number in the network. This can be explained by the decrease of the number of $Q_1$ and $Q_2$ in the network, consequently the competition on the best links reduces.

## IV. CONCLUSION AND FUTURE WORK

In this paper, we propose a bandwidth allocation mechanism for layered streaming in P2P network that allocates appropriate bandwidth to the appropriate peers while ensuring a minimum quality level to all peers. Each upstream peer organizes a set of auctions to sell its bandwidth, an auction for each layer, starting by the lower layers. In this manner the lower layers are transmitted via the best links, consequently increasing the system throughput.

To study the effectiveness of the proposed mechanism, we performed simulations and compared it with another existing system. We studied different metrics that are essential in determining the performance of our proposed mechanism. The results demonstrate the optimality and the effectiveness of our solution.